\documentclass{aa}  
\usepackage[varg]{txfonts}
\usepackage{graphicx}
\usepackage{amssymb,amsmath}
\usepackage{booktabs,multirow}
\usepackage{bm}
\usepackage{natbib}
\bibpunct{(}{)}{;}{a}{}{,}

\usepackage{color}

\begin{document}

\title{Relative magnetic field line helicity}

\author{K. Moraitis$^1$ \and E. Pariat$^1$ \and G. Valori$^2$ \and K. Dalmasse$^3$}
\institute{LESIA, Observatoire de Paris, Universit\'{e} PSL, CNRS, Sorbonne Universit\'{e}, Univ. Paris Diderot, Sorbonne Paris Cit\'{e}, 5 place Jules Janssen, 92195 Meudon, France \and Mullard Space Science Laboratory, University College London, Holmbury St. Mary, Dorking, Surrey, RH5 6NT, UK \and IRAP, Universit\'{e} de Toulouse, CNRS, CNES, UPS, 31028 Toulouse, France}

\date{Received ... / Accepted ...}

\abstract{Magnetic helicity is an important quantity in studies of magnetized plasmas as it provides a measure of the geometrical complexity of the magnetic field in a given volume. A more detailed description of the spatial distribution of magnetic helicity is given by the field line helicity that expresses the amount of helicity associated to individual field lines, rather than in the full analysed volume.}{Magnetic helicity is not a gauge-invariant quantity in general, unless it is computed with respect to a reference field, yielding the so-called relative magnetic helicity. The field line helicity corresponding to the relative magnetic helicity has only been examined under specific conditions so far. This work aims to define the field line helicity corresponding to relative magnetic helicity in the most general way. In addition to its general form, we provide the expression for the relative magnetic field line helicity in a few commonly used gauges, and reproduce known results as a limit of our general formulation.}{By starting from the definition of relative magnetic helicity, we derive the corresponding field line helicity, and we note the assumptions it is based on.}{We check that the developed quantity reproduces relative magnetic helicity by using three different numerical simulations. For these cases we also show the morphology of field line helicity in the volume, and on the photospheric plane. As an application to solar situations, we compare the morphology of field line helicity on the photosphere with that of the connectivity-based helicity flux density in two reconstructions of an active region's magnetic field. We discuss how the derived relative magnetic field line helicity has a wide range of applications, notably in solar physics and magnetic reconnection studies.}{}


\titlerunning{Relative magnetic field line helicity}
\authorrunning{Moraitis et al.}

\maketitle

%%%%%%%%%%%%%%%%%%%%%%%%%%%%%%%%%%%%%%%%%%%%%%%%%%%%%%%%%%%%%
\section{Introduction}
\label{sect:introduction}

Magnetic helicity is an important quantity in studies of the Sun, as well as, in any other environment where magnetohydrodynamics (MHD) is the appropriate description. This is because magnetic helicity is a conserved quantity in ideal MHD \citep{woltjer58}, and is approximately conserved even in non-ideal conditions, i.e., in the presence of reconnection (\citealt{taylor74}, and more recently, \citealt{pariat15}). 

Magnetic helicity quantifies the complexity of a magnetic field, and relates to the twist and braiding of the magnetic field lines. It is clasically defined as $H_m=\int_V \mathbf{A}\cdot \mathbf{B}\,{\rm d}V$, where $\mathbf{B}$ is the magnetic field in the volume $V$, and $\mathbf{A}$ the corresponding vector potential. It therefore depends on the gauge of $\mathbf{A}$ unless it is computed in a volume bounded by a magnetic flux surface. This condition can be too restrictive in many cases, especially when the considered volume is constrained by observational limitations as in solar observations. In order to remove this limitation, the helicity can be defined with respect to a reference magnetic field. The first such relative magnetic helicity was introduced by \citet{BergerF84}, and a gauge-independent version of it was later defined by \citet{fa85}. In this definition of relative magnetic helicity, the obtained helicity values do not depend on the gauges of the vector potentials of the original and reference fields as long as the normal components of the two fields coincide on the boundary of the volume. 

The importance of relative magnetic helicity (RMH) has lead to the development of many computational methods in the recent years. These include finite-volume methods in Cartesian coordinates \citep{thalmann11,val12,yang13,moraitis14}, finite-volume methods in spherical coordinates \citep{moraitis18}, and methods relying on other definitions of helicity \citep{pariat05,geo12a}. A complete description and intercomparison of the Cartesian methods can be found in \citet{valori16}.

Relative magnetic helicity is a global quantity that characterizes the magnetic field. There are cases however where it is desirable to be able to describe helicity more locally. This can be done with field line helicity that was introduced by \citet{antiochos87}, and \citet{berger88}. Field line helicity (FLH) of a magnetic field is simply the line integral of the vector potential (that produces the magnetic field) along a field line. The usefulness of FLH stems from the fact that its flux-weighted integral along the boundary is equal to the relative magnetic helicity. FLH can be considered as a function of the field lines, and, due to the arbitrariness of the gauge choice, has no physical meaning unless the field line is closed, in which case it expresses the flux of the magnetic field through the surface bounded by the closed field line. Alternatively, FLH can attain physical meaning under specific gauges for the vector potentials.

Field line helicity has been employed in various solar applications. As an example, \citet{russel15} used FLH to trace the connectivity changes caused by reconnection in a magnetic field confined between two planes, simulating coronal loops. In another example, \citet{yeates16} have applied FLH to the global magnetic field of the Sun in order to study the distribution of helicity in the solar corona. A more theoretical study of the properties of FLH can be found in \citet{aly18}. In most of these applications a special gauge for the vector potentials was used to simplify calculations, which lead to inaccurate helicity budgets when the flux-weighted integral of FLH along the boundary was taken. This has very recently been fixed by \citet{yeates18}, where a relative FLH was introduced that reproduced relative magnetic helicity to a high degree when integrated over the boundary. Still, this relative FLH is in the same constrained gauge used previously. One of the purposes of this paper is to find an expression for field line helicity that accurately reproduces relative magnetic helicity, and that, at the same time, makes no assumption about the adopted gauge.

Apart from field line helicity there are other methods that employ the magnetic connectivity to compute quantities related to magnetic helicity. An example is the twist number method \citep{guo17}, where the twist component of helicity is computed in configurations consisting of twisted magnetic flux ropes. In an alternative method that approximates helicity through its twist, \citet{malanushenko11} compute RMH by matching the shapes of coronal loops with field lines of a linear force-free field.

Another method that uses the magnetic connectivity to infer a helicity-related quantity is the connectivity-based helicity flux density method \citep{pariat05,dalmasse14}. In this, the rate of helicity change in the volume (${\rm d}H_m/{\rm d}t$) is deduced from the flux-weighted integral of a quantity ${\rm d}h_\Phi/{\rm d}t$ along the boundary of the volume, which can therefore be considered as the helicity flux density per unit of magnetic flux. Moreover, this flux density is inferred from the connectivity of the magnetic field in a similar manner as FLH.

In this paper we define a relative magnetic field line helicity (RMFLH) whose gauge is unrestricted, and investigate its relation with the connectivity-based helicity flux density. In Sect.~\ref{sect:rmflh} we define RMFLH, and also express it in various simplifying gauges. In Sect.~\ref{sect:data} we describe the magnetic field data that we use for testing our derivation. In Sect.~\ref{sect:application} we first validate the derived field line helicity method, and then we apply and compare it to a similar method that uses the helicity flux density. Finally, in Sect.~\ref{sect:discussion} we summarize and discuss the results of the paper.

%%%%%%%%%%%%%%%%%%%%%%%%%%%%%%%%%%%%%%%%%%%%%%%%%%%%%%%%%%%%%
\section{Relative magnetic field line helicity}
\label{sect:rmflh}

\subsection{Definition}
\label{sect:rmflh1}

Relative magnetic helicity of a three-dimensional (3D) magnetic field $\mathbf{B}$ in a simply-connected, finite volume $V\subset\mathbb{R}^3$ with boundary $\partial V$, is given by the \citet{fa85} formula
\begin{equation}
H_r=\int_V (\mathbf{A}+\mathbf{A}_\mathrm{p})\cdot (\mathbf{B}-\mathbf{B}_\mathrm{p})\,{\rm d}V.
\label{helr}
\end{equation}
Here, $\mathbf{B}_\mathrm{p}$ is a potential magnetic field that is used as reference, and $\mathbf{A}$, $\mathbf{A}_\mathrm{p}$ are the vector potentials of the fields $\mathbf{B}$, $\mathbf{B}_\mathrm{p}$, respectively. Relative magnetic helicity does not depend on the gauges of $\mathbf{A}$ and $\mathbf{A}_\mathrm{p}$ as long as the condition
\begin{equation}
\left. \hat{n}\cdot \mathbf{B} \right|_{\partial V}=\left. \hat{n}\cdot \mathbf{B}_\mathrm{p} \right|_{\partial V}
\label{helc}
\end{equation}
holds \citep[see, e.g.,][]{val12}, where $\hat{n}$ denotes the outward-pointing unit normal of $\partial V$. The relative magnetic helicity defined in Eq.~(\ref{helr}) can be decomposed into the sum of two gauge-independent helicities, one involving only the current-carrying part of the magnetic field, $\mathbf{B}_\mathrm{j}=\mathbf{B}-\mathbf{B}_\mathrm{p}$, namely
\begin{equation}
H_j=\int_V (\mathbf{A}-\mathbf{A}_\mathrm{p})\cdot (\mathbf{B}-\mathbf{B}_\mathrm{p})\,{\rm d}V,
\label{helj}
\end{equation}
and one also involving the potential magnetic field
\begin{equation}
H_{pj}=2\int_V \mathbf{A}_\mathrm{p}\cdot (\mathbf{B}-\mathbf{B}_\mathrm{p})\,{\rm d}V.
\label{helpj}
\end{equation}
These components are referred to as the non-potential and volume-threading helicities in \citet{linan18}, respectively.

In order to derive a general expression for the field line helicity we start from relative magnetic helicity of Eq.~(\ref{helr}), and decompose it as
\begin{equation}
H_r=\int_V\, (\mathbf{A}+\mathbf{A}_\mathrm{p})\cdot \mathbf{B}\,{\rm d}V - \int_V\, (\mathbf{A}+\mathbf{A}_\mathrm{p})\cdot \mathbf{B}_\mathrm{p}\,{\rm d}V,
\label{helc2}
\end{equation}
where the individual terms on the right-hand side are now gauge-dependent. The volume integrations can be performed along the field lines of the respective magnetic field if we split the volume elements according to ${\rm d}V={\rm d}S{\rm d}l={\rm d}\mathbf{S}\cdot {\rm d}\bm{l}$ (with both ${\rm d}\mathbf{S}$, ${\rm d}\bm{l}$ pointing along the field lines) in the first integral, and ${\rm d}V={\rm d}S{\rm d}l_\mathrm{p}={\rm d}\mathbf{S}\cdot {\rm d}\bm{l}_\mathrm{p}$ in the second, and rearrange the dot products inside the integrals. Equation (\ref{helc2}) then reads
\begin{multline}
H_r=\int_{\partial V^+} \left| \hat{n}\cdot\mathbf{B} \right| \left( \int_{\alpha_+}^{\alpha_-}\,(\mathbf{A}+\mathbf{A}_\mathrm{p}) \cdot {\rm d}\bm{l} \right)\,{\rm d}S - \\
\int_{\partial V^+} \left| \hat{n}\cdot\mathbf{B}_\mathrm{p} \right| \left( \int_{\alpha_{p+}}^{\alpha_{p-}}\,(\mathbf{A}+\mathbf{A}_\mathrm{p}) \cdot {\rm d}\bm{l}_\mathrm{p} \right)\,{\rm d}S ,
\label{helc3}
\end{multline}
where $\partial V^+$ ($\partial V^-$) denotes the part of the boundary where magnetic flux enters into (leaves) the volume, defined through $\partial V^\pm=\{\mathbf{x}\in \partial V: \hat{n}\cdot \mathbf{B}(\mathbf{x})\lessgtr 0\}$. The footpoints of the field lines of $\mathbf{B}$ ($\mathbf{B}_\mathrm{p}$) are denoted as $\alpha_+\in \partial V^+$, $\alpha_-\in \partial V^-$ ($\alpha_{p+}\in \partial V^+$, $\alpha_{p-}\in \partial V^-$). The only assumption made in going from Eq.~(\ref{helc2}) to (\ref{helc3}) is that each field line can be labelled with its infinitesimal flux, or in other words, that an infinitesimal flux tube can surround each field line. This assumption can always be fulfilled in regions of sufficiently smooth magnetic field.

The integrations in Eq.~(\ref{helc3}) are taken along $\partial V^+$ so that each field line is counted only once, but equivalently, they can be performed along $\partial V^-$, or even along the whole boundary, $\partial V$, but in that case the resulting helicity should be halved. Since by construction the potential field has the same normal components along the boundaries with the given magnetic field, we can choose to start from the same footpoint in the two integrals, $\alpha_{p+}=\alpha_+$, so that $\left| \hat{n}\cdot\mathbf{B} \right|=\left| \hat{n}\cdot\mathbf{B}_\mathrm{p} \right|$, yielding
\begin{equation}
H_r=\int_{\partial V^+} \left| \hat{n}\cdot\mathbf{B} \right| \left( \int_{\alpha_+}^{\alpha_-}\,(\mathbf{A}+\mathbf{A}_\mathrm{p}) \cdot {\rm d}\bm{l} - \int_{\alpha_+}^{\alpha_{p-}}\,(\mathbf{A}+\mathbf{A}_\mathrm{p}) \cdot {\rm d}\bm{l}_\mathrm{p} \right)\,{\rm d}S.
\label{flh}
\end{equation}
By defining the relative magnetic field line helicity as
\begin{equation}
\mathcal{A}_r^+= \int_{\alpha_+}^{\alpha_-}\,(\mathbf{A}+\mathbf{A}_\mathrm{p}) \cdot {\rm d}\bm{l} - \int_{\alpha_+}^{\alpha_{p-}}\,(\mathbf{A}+\mathbf{A}_\mathrm{p}) \cdot {\rm d}\bm{l}_\mathrm{p},
\label{flhdef}
\end{equation}
it follows that $H_r=\int_{\partial V^+} \mathcal{A}_r^+\,{\rm d}\Phi$ with ${\rm d}\Phi=\left| \hat{n}\cdot\mathbf{B} \right|\,{\rm d}S$ the elementary magnetic flux on the boundary. A similar expression results when we consider as starting points the field lines of $\partial V^-$
\begin{equation}
\mathcal{A}_r^-= \int_{\alpha_+}^{\alpha_-}\,(\mathbf{A}+\mathbf{A}_\mathrm{p}) \cdot {\rm d}\bm{l} - \int_{\alpha_{p+}}^{\alpha_{-}}\,(\mathbf{A}+\mathbf{A}_\mathrm{p}) \cdot {\rm d}\bm{l}_\mathrm{p},
\label{flhdefm}
\end{equation}
where the common footpoint is now $\alpha_{p-}=\alpha_{-}$. When the whole boundary is considered, the respective RMFLH is given by the average of the latter two expressions, namely
%\begin{multline}
%\mathcal{A}_r^0=\frac{1}{2}\left( \mathcal{A}_r^+ + \mathcal{A}_r^-\right)  \int_{\alpha_+}^{\alpha_-}\,(\mathbf{A}+\mathbf{A}_\mathrm{p}) \cdot {\rm d}\bm{l} - \frac{1}{2}\left( \int_{\alpha_+}^{\alpha_{p-}}\,(\mathbf{A}+\mathbf{A}_\mathrm{p}) \cdot {\rm d}\bm{l}_\mathrm{p} \right. +\\
%\left. \int_{\alpha_{p+}}^{\alpha_{-}}\,(\mathbf{A}+\mathbf{A}_\mathrm{p}) \cdot {\rm d}\bm{l}_\mathrm{p} \right).
%\label{flhdef0}
%\end{multline}
\begin{equation}
\mathcal{A}_r^0=\frac{1}{2}\left( \mathcal{A}_r^+ + \mathcal{A}_r^-\right).
\label{flhdef0}
\end{equation}
In all cases the RMFLH is expressed relative to the same quantity integrated along the field lines of the potential magnetic field, hence the name relative magnetic field line helicity seems appropriate. It also follows that for a potential magnetic field all RMFLH expressions, Eqs.~(\ref{flhdef}), (\ref{flhdefm}), and (\ref{flhdef0}), vanish by definition. Inevitably, the RMFLH is a function of not only the footpoints of $\mathbf{B}$, but also of $\mathbf{B}_\mathrm{p}$. Expressed differently, RMFLH depends on the way that both fields connect positive to negative footpoints.

The relative magnetic helicity can be interpreted as the helicity budget of the composing field line helicities in the whole volume, and can be expressed in all cases as
\begin{equation}
H_r=\int_{\partial V^s} \mathcal{A}_r^s\,{\rm d}\Phi,
\label{flhhel}
\end{equation}
where $s$ stands for the characters, `+', `-', or `0', and we defined for reasons of uniformity, $\partial V^0=\partial V$. This relation shows that the RMFLH can be considered as the RMH density per unit magnetic flux. Note also that, differently from the direct computation of RMH as a volume integral, its computation as a budget of RMFLH does not include the contribution from field lines that do not connect with the boundary (if any) but rather close within the volume. The RMFLH can always be computed from one of the Eqs.~(\ref{flhdef})--(\ref{flhdef0}), but its flux-weighted integral given by Eq.~(\ref{flhhel}) will be equal to the RMH only if all field lines are connected to the boundary at both ends, i.e., when no closed and/or ergodic field lines are present in the volume. We adopt this assumption henceforth.

In a similar manner one can also define the relative magnetic field line helicity of the components $H_j$ and $H_{pj}$ of helicity by replacing $\mathbf{A}+\mathbf{A}_\mathrm{p}$ in Eqs.~(\ref{flhdef})--(\ref{flhdef0}) with $\mathbf{A}-\mathbf{A}_\mathrm{p}$, and $2\mathbf{A}_\mathrm{p}$, respectively. As an example, the relative magnetic field line helicity of the non-potential component of helicity with field lines starting from the positive polarity is
\begin{equation}
\mathcal{A}_j^+= \int_{\alpha_+}^{\alpha_-}\,(\mathbf{A}-\mathbf{A}_\mathrm{p}) \cdot {\rm d}\bm{l} - \int_{\alpha_+}^{\alpha_{p-}}\,(\mathbf{A}-\mathbf{A}_\mathrm{p}) \cdot {\rm d}\bm{l}_\mathrm{p}
\label{flhdefj}
\end{equation}
Of course, all these RMFLHs are gauge-dependent, but their flux-weighted integrals along the (whole) boundary recover the respective gauge-independent helicity component, i.e., for Eq.~(\ref{flhdefj}) the budget $H_j=\int_{\partial V^+} \mathcal{A}_j^+\,{\rm d}\Phi$ is indeed gauge-invariant.

\subsection{Specific gauge choices}
\label{sect:rmflh2}

As with any FLH defined on open field lines, all RMFLH expressions depend on the gauges of the vector potentials. The only gauge independent FLH is when field lines are closed, in which case the FLH is the magnetic flux through the surface defined by the closed field line. For the RMFLH, by setting $\mathbf{A}'=\mathbf{A}+\nabla \psi_1$ and $\mathbf{A}'_\mathrm{p}=\mathbf{A}_\mathrm{p}+\nabla \psi_2$ in Eq.~(\ref{flhdef}) for example, it can easily be seen that $\mathcal{A}^{+'}_r=\mathcal{A}^+_r+(\psi_1+\psi_2)(\alpha_{-})-(\psi_1+\psi_2)(\alpha_{p-})$. It would be thus useful to exploit this gauge freedom in order to derive expressions for the RMFLH that are a function of the footpoints of only one magnetic field (and preferably of $\mathbf{B}$) instead of two, and, if possible, without restricting the gauge of $\mathbf{A}$ in doing so. Here we give a non-exhaustive list of choices for the gauges that simplify the resulting RMFLH.

\subsubsection{Potential-free gauge}

An obvious way to have a RMFLH that depends only on the footpoints of $\mathbf{B}$ is by making the second term in Eqs.~(\ref{flhdef}), (\ref{flhdefm}) to vanish. By going back in the derivation to Eq.~(\ref{helc2}), and replacing in the second term the potential field with the identity $\mathbf{B}_\mathrm{p}=\nabla\Psi$, we have
\begin{multline*}
H_r=\int_V\, (\mathbf{A}+\mathbf{A}_\mathrm{p})\cdot \mathbf{B}\,{\rm d}V - \int_V\, (\mathbf{A}+\mathbf{A}_\mathrm{p})\cdot \nabla\Psi\,{\rm d}V=\\
\int_V\, (\mathbf{A}+\mathbf{A}_\mathrm{p})\cdot \mathbf{B}\,{\rm d}V - \oint_{\partial V}\, \Psi(\mathbf{A}+\mathbf{A}_\mathrm{p})\cdot {\rm d}\mathbf{S}+\int_V\, \Psi \nabla\cdot(\mathbf{A}+\mathbf{A}_\mathrm{p}) \,{\rm d}V.
\label{helc2b}
\end{multline*}
The last two terms can be eliminated with the gauge combination
\begin{equation}
\begin{cases}
\nabla\cdot\mathbf{A}_\mathrm{p}=-\nabla\cdot\mathbf{A}\\
\left. \hat{n}\cdot \mathbf{A}_\mathrm{p} \right|_{\partial V}=-\left. \hat{n}\cdot \mathbf{A} \right|_{\partial V}
\end{cases}
\end{equation}
The respective field line helicity is then
\begin{equation}
\mathcal{A}_r^A= \int_{\alpha_+}^{\alpha_-}\,(\mathbf{A}+\mathbf{A}_\mathrm{p}) \cdot {\rm d}\bm{l}.
\label{flhgaug1}
\end{equation}
It depends only on the footpoints of $\mathbf{B}$, and can be obtained with an arbitrary gauge for $\mathbf{A}$, only the gauge of $\mathbf{A}_\mathrm{p}$ is restricted.

\subsubsection{Berger \& Field gauge}

Another possibility for the gauge is given by the relation
\begin{equation}
\left. \hat{n}\cdot\left(\mathbf{A}\times \mathbf{A}_\mathrm{p}\right) \right|_{\partial V}=0,
\label{eq:gaugeb}
\end{equation}
which involves the tangential components of the two vector potentials on the boundary, as can be seen from the sufficient condition for its validity
\begin{equation}
\left. \hat{n}\times\mathbf{A}_\mathrm{p} \right|_{\partial V} = \left. \hat{n}\times\mathbf{A} \right|_{\partial V}.
\label{eq:gaugeb2}
\end{equation}
This gauge can be fulfilled with the tangential components of $\mathbf{A}_\mathrm{p}$ being determined from those of $\mathbf{A}$, while the gauge of the latter is unrestricted. The gauge condition given by Eq.~(\ref{eq:gaugeb2}) was used in the original definition of relative magnetic helicity by \citet{BergerF84}, where Eq.~(\ref{helc}) then reads $H_r=\int_V \mathbf{A}\cdot \mathbf{B}\,{\rm d}V-\int_V \mathbf{A}_\mathrm{p}\cdot \mathbf{B}_\mathrm{p}\,{\rm d}V$. This gauge is thoroughly examined in \citet{yeates18} where it is used to define the following relative field line helicity
\begin{equation}
\mathcal{A}_r^{\rm YP,+}= \int_{\alpha_+}^{\alpha_-}\,\mathbf{A} \cdot {\rm d}\bm{l} - \int_{\alpha_+}^{\alpha_{p-}}\,\mathbf{A}_\mathrm{p} \cdot {\rm d}\bm{l}_\mathrm{p},
\label{flhyp}
\end{equation}
written here in our notation and for the positive polarity boundary. This relative FLH is constructed similarly to the RMFLH of this paper since both quantities are differences of FLHs along the field lines of $\mathbf{B}$ and $\mathbf{B}_\mathrm{p}$. Unlike the RMFLH given by Eq.~(\ref{flhdef}) however, it employs a specific gauge condition, although \citet{yeates18} provide a recipe to impose this gauge starting from a general one.

An alternative form for the relative FLH of \citet{yeates18} can be obtained when helicity is written as $H_r=\int_V (\mathbf{A}-\mathbf{A}_\mathrm{p})\cdot (\mathbf{B}+\mathbf{B}_\mathrm{p})\,{\rm d}V$ in the same gauge \citep{BergerF84}. By splitting the volume element along and across the field lines of the magnetic field $\mathbf{B}_\mathrm{q}=\mathbf{B}+\mathbf{B}_\mathrm{p}$ according to ${\rm d}V={\rm d}\mathbf{S}\cdot {\rm d}\bm{l}_\mathrm{q}$, we can define the respective field line helicity 
\begin{equation}
\mathcal{A}_r^B= 2\int_{\alpha_{q+}}^{\alpha_{q-}}\,(\mathbf{A}-\mathbf{A}_\mathrm{p}) \cdot {\rm d}\bm{l}_\mathrm{q}.
\label{flhgaug2}
\end{equation}
With respect to \citet{yeates18}, this RMFLH depends on the footpoints ($\alpha_{q+},\alpha_{q-}$) of the single magnetic field $\mathbf{B}_\mathrm{q}$, and therefore requires to trace field lines for this magnetic field only. The factor of 2 comes from the flux of the respective flux tubes, ${\rm d}\Phi_\mathrm{q}=2{\rm d}\Phi$, on the boundary.

\subsubsection{Original Berger gauge}

An even more restricting gauge combination is the one originally used in \citet{berger88}, and more recently in \citet{yeates16}. This consists of the relations 
\begin{equation}
\begin{cases}
\nabla\cdot\mathbf{A}_\mathrm{p}=0\\
\left. \hat{n}\cdot \mathbf{A}_\mathrm{p} \right|_{\partial V}=0\\
\left. \hat{n}\cdot\left(\mathbf{A}\times \mathbf{A}_\mathrm{p}\right) \right|_{\partial V}=0
\end{cases}
\label{gaugeC}
\end{equation}
and it also involves the gauge of $\mathbf{A}$, not just $\mathbf{A}_\mathrm{p}$'s. Relative magnetic helicity from Eq.~(\ref{helc}) simplifies then to $H_r=\int_V \mathbf{A}\cdot \mathbf{B}\,{\rm d}V$ \citep[see also][]{val12}. By splitting the volume element along and across the field lines of $\mathbf{B}$, we derive the respective field line helicity 
\begin{equation}
\mathcal{A}_r^C= \int_{\alpha_+}^{\alpha_-}\,\mathbf{A} \cdot {\rm d}\bm{l}.
\label{flhgaug3}
\end{equation}
This RMFLH involves only the footpoints of $\mathbf{B}$. Although the form of RMFLH in this gauge is very simple, it is difficult to use in practical implementations because of all the restrictions of Eq.~(\ref{gaugeC}).

%%%%%%%%%%%%%%%%%%%%%%%%%%%%%%%%%%%%%%%%%%%%%%%%%%%%%%%%%%%%%
\section{Magnetic-field data}
\label{sect:data}

To check that the definition of relative magnetic field line helicity, given in Sect.~\ref{sect:rmflh}, is consistent, and that it is implemented correctly, we need to compare the values of helicity obtained with the RMFLH method with those of a standard volume method. We use three MHD simulations for this comparison, so that to be able to also follow the evolution of the RMH. 

Additionally, we want to examine the relation between the helicity flux density method of \cite{dalmasse14} with the RMFLH one. For this we use two extrapolations of the 3D magnetic field of an observed active region (AR). All magnetic field data, from simulations or from extrapolations, are described below.

\subsection{MHD data}
\label{sect:data1}

The first two MHD datasets that we use are from the non-eruptive \citep{leake13}, and eruptive \citep{leake14} flux-emergence simulations that were used and described in \citet{valori16}, and also in \citet{pariat17} and in \citet{linan18}. The two simulations involve the emergence of the same twisted magnetic flux rope in the same stratified solar atmosphere, and differ only in the strength and orientation of the initial, background, coronal magnetic field. In the non-eruptive case the initial field is parallel to the emerging flux rope, while in the eruptive case it is anti-parallel (and smaller than the non-eruptive one), thus favoring magnetic reconnection between emerging and background fields. 

\begin{figure*}[h]
\centering
\includegraphics[width=0.9\textwidth]{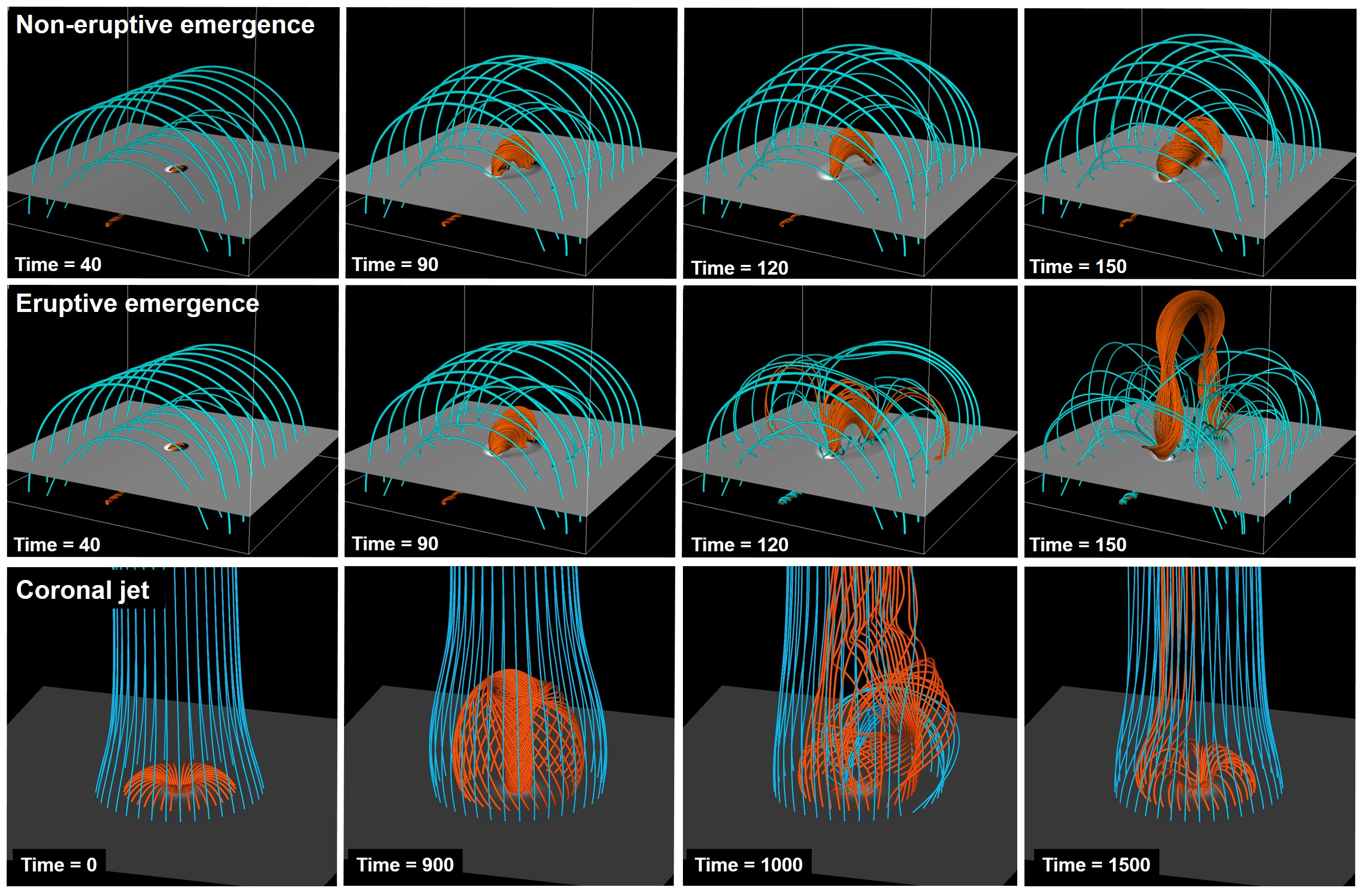}
\caption{Evolution of the magnetic field morphology for the three MHD simulations described in Sect.~\ref{sect:data1}. Shown are the $B_z$-distribution at the photospheric-like plane, and selected magnetic field lines of the current-carrying structure (with orange) and of the surrounding field (with blue).}
\label{datafig1}
\end{figure*}

The magnetic field in both simulations is interpolated from the original stretched grid to a uniform one, that spans the volume $[-100,100]\times[-100,100]\times[0.36,150]$ in the non-dimensional units of the MHD code. The resulting mesh consists of 233$\times$233$\times$174 grid points, with a pixel size of 0.859 in all directions. The simulations last from $t=0$ to $t=200$. In the eruptive case an eruption occurs around $t=120$. The morphology of the magnetic field and the way it changes during the simulation is shown in the top two panels of Fig.~\ref{datafig1} for both simulations. There, one can see the emerging flux rope (in orange), and the overlying arcade field (in blue), and the different behavior of their interaction in the two simulations.

The third dataset that we used is based on the simulation of \citet{pad09} for the formation of a coronal jet, and is described in \citet{pariat15} and in \citet{linan18}. In the simulation, there is initially uniform coronal plasma around an axisymmetric null point, that is created by embedding a vertical dipole below the simulation domain, and adding a uniform volume vertical magnetic field of opposite direction in the domain. As a result, the initial field in the volume is split into two connectivity domains, one closed and one open. Photospheric twisting motions eventually destabilize the configuration and magnetic reconnection between closed and open field lines occurs, with the final outcome the generation of a jet.

The magnetic field occupies the volume $[-6,6]\times[-6,6]\times[0,12]$ (in non-dimensional units), and is discretized with 129$\times$129$\times$129 grid points, and a pixel size of 0.0938 in all directions. The simulation lasts from $t=0$ to $t=1600$, with an accumulation phase for $t\in[0,700]$, and a dynamical phase for $t\in[700,1600]$ that is sampled more frequently, while the non-ideal effects start to appear around $t\approx 920$. The evolution of the magnetic field morphology for this simulation is shown in the bottom panel of Fig.~\ref{datafig1}.

\subsection{Extrapolation data}
\label{sect:data2}

To investigate the relation of field line helicity with the connectivity-based helicity flux density, ${\rm d}h_\Phi/{\rm d}t$, we use the same 3D magnetic-field data as in \citet{dalmasse18}. These comprise a potential and a non-linear force-free (NLFF) reconstruction of the 3D magnetic field of NOAA AR 11158. Both methods start from the same vector magnetogram, which is the average of the SDO/HMI ones at 06:22 UT and 06:34 UT of 14 February 2011, rebinned by a factor of two, to the resolution $1\arcsec$ per pixel. 

The first reconstruction produces a potential magnetic field through the scalar potential of \citet{schmidt64}. This method uses only the normal component of the photospheric magnetic field. The scalar potential is obtained from the numerical solution of Laplace's equation with the Dirichlet boundary conditions along the six boundaries of the volume that are given by the actual Schmidt solution. The resulting potential field occupies the volume $204\,{\rm Mm}\times 200\,{\rm Mm}\times 144\,{\rm Mm}$, which is discretized by $284\times 277 \times 200$ grid points.

\begin{figure}[h]
\centering
\includegraphics[width=0.42\textwidth]{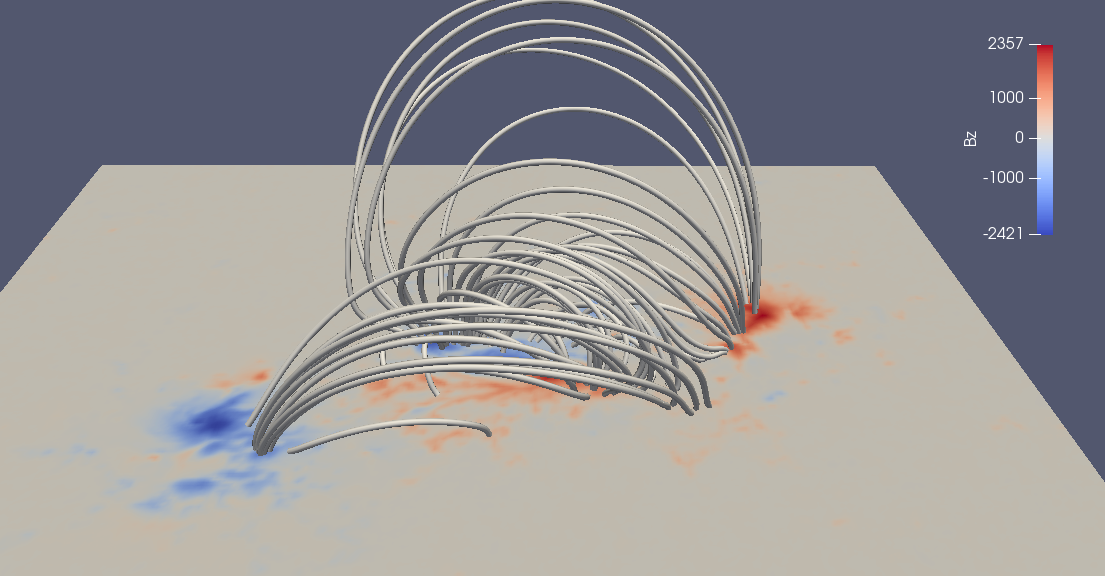}\\
\includegraphics[width=0.42\textwidth]{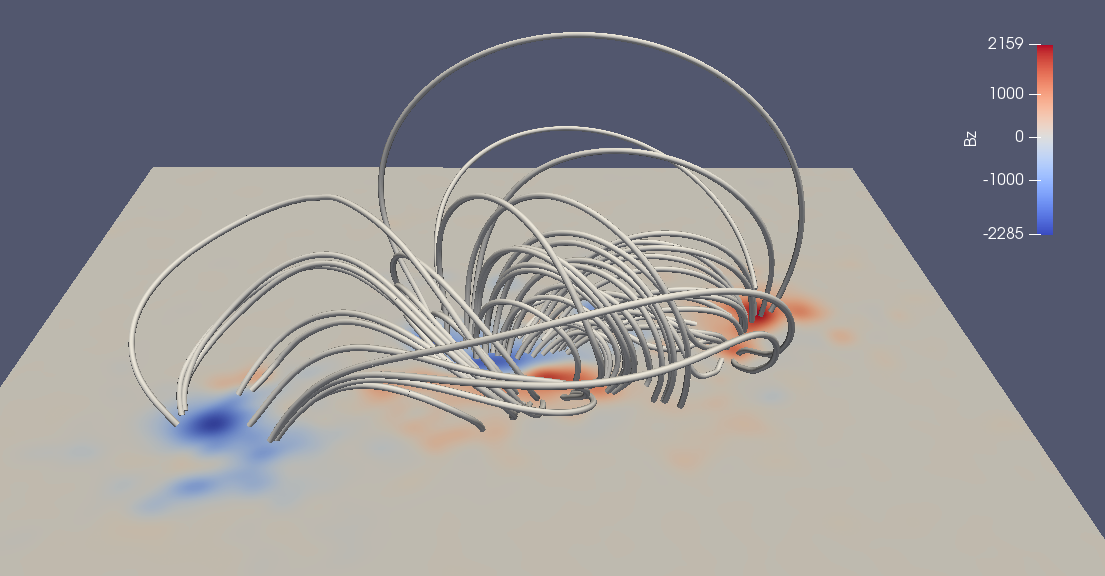}
\caption{Morphology of the 3D magnetic field of NOAA AR 11158 on 06:28 UT of 14 February 2011, for the potential (top) and the NLFF (bottom) extrapolation, as described in Sect.~\ref{sect:data2}. The same seed points for drawing the field lines are used in both cases.}
\label{datafig2}
\end{figure}

The second reconstruction is a NLFF one that uses a weighted optimization method \citep{wieg04}, after pre-processing the horizontal components of the field on the photosphere so that it becomes more compatible with the force-free assumption, and additionally smoothing of the original vector magnetogram. The resulting NLFF field occupies the volume $185\,{\rm Mm}\times 185\,{\rm Mm}\times 144\,{\rm Mm}$, which is discretized by $256\times 256 \times 200$ grid points. The full details on each method can be found in \citet{dalmasse18} and references therein. The morphology of the magnetic field for the two extrapolations is shown in Fig.~\ref{datafig2}.

%%%%%%%%%%%%%%%%%%%%%%%%%%%%%%%%%%%%%%%%%%%%%%%%%%%%%%%%%%%%%
\section{Applications}
\label{sect:application}

In this section we present the results from the application of the field line helicity method of Sect.~\ref{sect:rmflh} to the magnetic field data of Sect.~\ref{sect:data}. Before that, we describe a few details of the implementation of the method.

\subsection{Implementation details of the RMFLH method}
\label{sect:application1}

The calculation of relative magnetic helicity with the help of relative magnetic field line helicity is an instantaneous integration method that needs as input the 3D magnetic field in the volume under study. It can thus be considered as a finite-volume method according to the classification of \citet{valori16}, and consequently it can be computed simultaneously with any method that employs the definition of Eq.~(\ref{helr}), although RMFLH needs additionally the computation of the field lines. 

Since all the test data of Sect.~\ref{sect:data} are in the Cartesian domain, we follow for each snapshot of the given 3D magnetic field the volume method described in \citet{moraitis14}. We first compute numerically the potential field that satisfies the condition of Eq.~(\ref{helc}) as a solution of Laplace's equation. From the two magnetic fields, $\mathbf{B}$ and $\mathbf{B}_\mathrm{p}$, we then compute their corresponding vector potentials by adopting the DeVore gauge, for which the vertical component is assumed to be null, $A_z=0$. This gauge was originally introduced by \citet{devore00} and was then modified by \citet{val12} for use in finite volumes, and it has the advantage of being very accurate and fast to compute. Our computation of RMFLH is thus performed in the most efficient way and it does not assume any of the specific gauge choices presented in Sect.~\ref{sect:rmflh2}. The quantity ($\mathbf{A}+\mathbf{A}_\mathrm{p}$) is then input to the field line integration part, along with the two magnetic fields.

The relative magnetic field line helicities $\mathcal{A}_r^+$ and $\mathcal{A}_r^-$ are computed from Eqs.~(\ref{flhdef}) and (\ref{flhdefm}) by calculating the two terms separately and then taking their difference. Each of the terms in these relations is obtained by an integration along the respective field lines, and so this method involves two different field line integrations, one for $\mathbf{B}$, and one for $\mathbf{B}_\mathrm{p}$. For these integrations we use as a starting point the code QSL Squasher \citep{qslsquasher} which is very fast and robust, and we modify it in three ways. First, we keep only the field-line integration part and omit any squashing factor calculations. Second, we modify the starting points of the field lines to be user-supplied instead of automatically determined by space-filling Hilbert curves. Finally, we augment the system of the three equations 
\begin{equation}
\frac{{\rm d}\bm{l}}{{\rm d}s}=\frac{\mathbf{B}}{B}
\end{equation}
that are solved by the code with one more, which describes the field line helicity change (${\rm d}h$) along a field line as
\begin{equation}
\frac{{\rm d}h}{{\rm d}s}=\frac{(\mathbf{A}+\mathbf{A}_\mathrm{p})\cdot\mathbf{B}}{B},
\end{equation}
with $s$ denoting the distance along the given field line.

For the starting points of the field lines (the footpoints) we consider separately the positive- and the negative-polarity boundary. Once the field lines are obtained, we compute either $\mathcal{A}_r^+$ or $\mathcal{A}_r^-$ 
depending on whether the field lines start from the positive or negative magnetic sign domain, using Eq.~(\ref{flhdef}) or Eq.~(\ref{flhdefm}), respectively, while for $\mathcal{A}_r^0$ we simply take the average of these expressions as in Eq.~(\ref{flhdef0}). All the results presented in the following use the grid points of the given magnetic field as the footpoints for the field lines. An alternative method which gives more weight to locations of higher magnetic field is discussed in the Appendix.

\subsection{Application to MHD data}
\label{sect:application2}

\begin{figure}[h]
\centering
\includegraphics[width=0.45\textwidth]{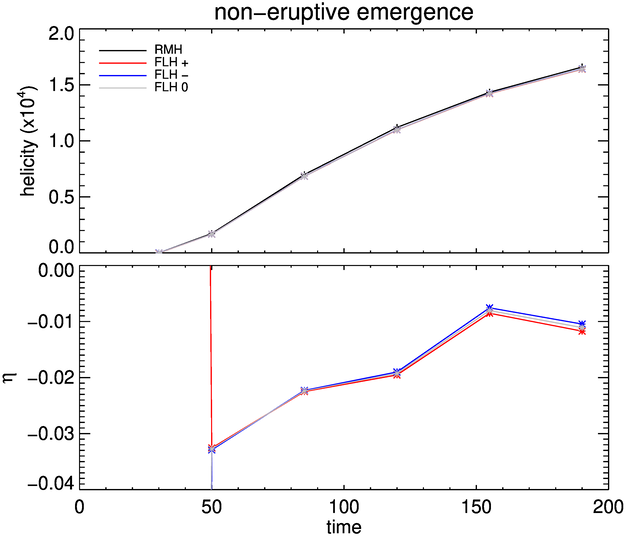}\\
\includegraphics[width=0.45\textwidth]{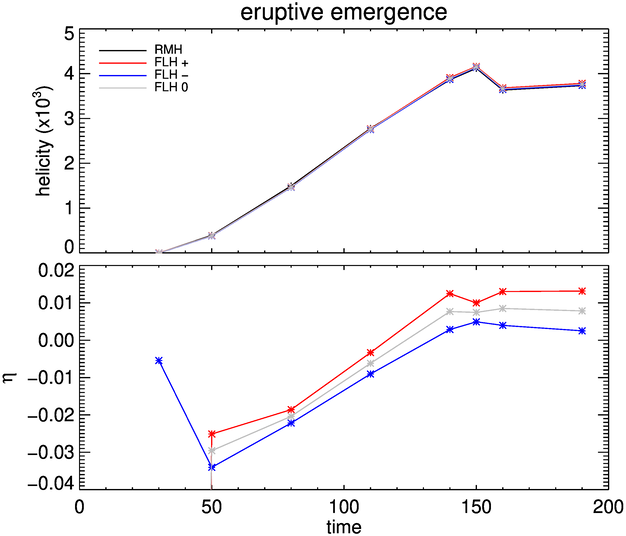}\\
\includegraphics[width=0.45\textwidth]{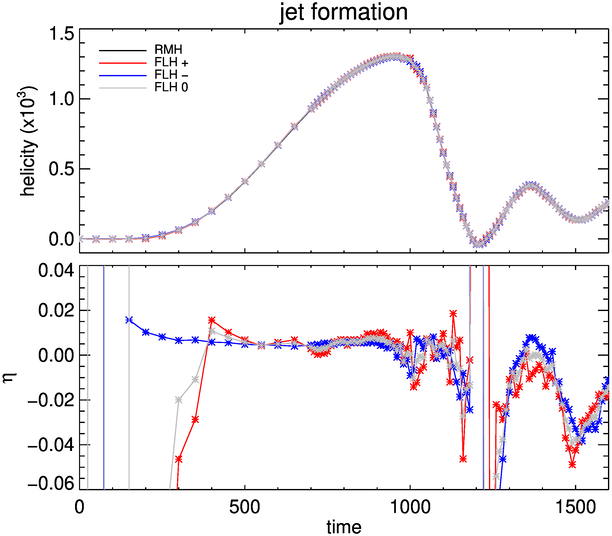}
\caption{Evolution of relative magnetic helicity as calculated by the volume method (with Eq.~(\ref{helr}), black curve) and by the field line method (with Eq.~(\ref{flhhel}), colored curves), and of their relative difference $\eta$, for the non-eruptive emergence (top), the eruptive emergence (middle), and the jet formation simulations (bottom). In all cases it is shown the RMFLH that starts from the positive footpoints (Eq.~(\ref{flhdef}), red curve), from the negative footpoints (Eq.~(\ref{flhdefm}), blue curve), and from both (Eq.~(\ref{flhdef0}), grey curve).}
\label{compfig0}
\end{figure}

The verification of the RMFLH-based helicity calculation method presented in Sect.~\ref{sect:rmflh} is done with the help of the three MHD simulations. For each snapshot of each simulation we compute the relative magnetic helicity from its definition, Eq.~(\ref{helr}), and also through the field line method, using Eq.~(\ref{flhhel}) and the RMFLH given either by Eqs.~(\ref{flhdef}), (\ref{flhdefm}), or (\ref{flhdef0}). The resulting evolution of the relative magnetic helicity with the two methods for all simulations is shown in Fig.~\ref{compfig0}. 

We notice that the agreement between the two methods is very good for all cases with the curves being practically indistinguishable. To look closer at it, we denote with $H_1$ the relative magnetic helicity budget calculated from Eq.~(\ref{helr}) and $H_2$ the one from Eq.~(\ref{flhhel}), and we define their relative difference as
\begin{equation}
\eta=\frac{H_2-H_1}{H_1}.
\end{equation}
The evolution of this relative difference for all MHD simulations is also shown in Fig.~\ref{compfig0}. Except from the snapshots at the start of the simulations and a few more where helicity is very low, all differences are within $\pm 5\%$. The computation of helicity with the RMFLH method thus gives results very similar with the volume method in all cases. Additionally, this is true irrespective of whether the starting footpoints belong to the positive or the negative polarity. These observations indicate that the equations derived in Sect.~\ref{sect:rmflh} are indeed correct, and also that they are properly implemented.

\begin{figure}[h]
\centering
\includegraphics[width=0.47\textwidth]{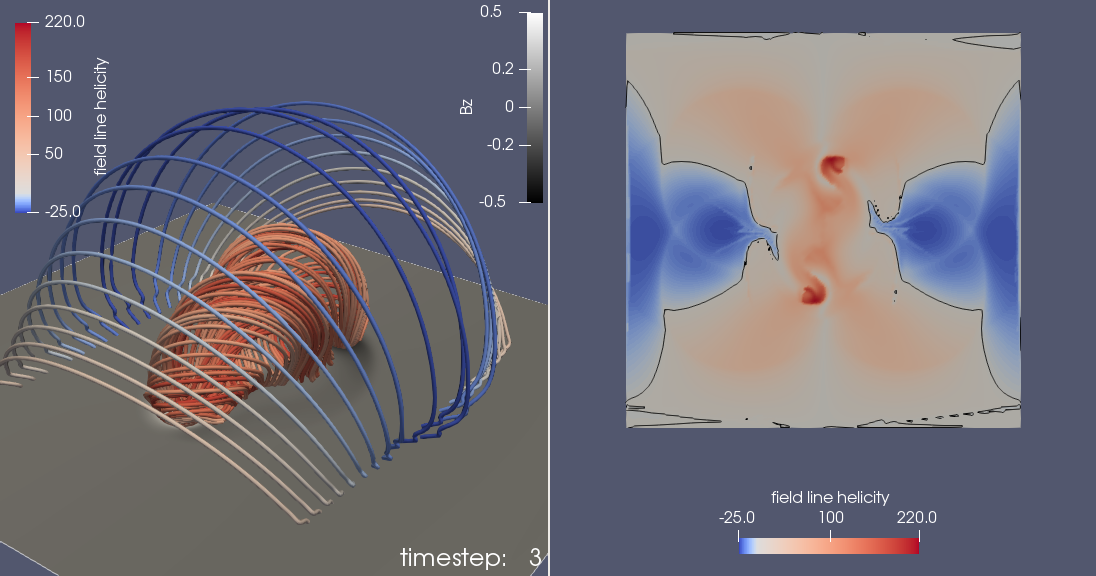}\\
\includegraphics[width=0.47\textwidth]{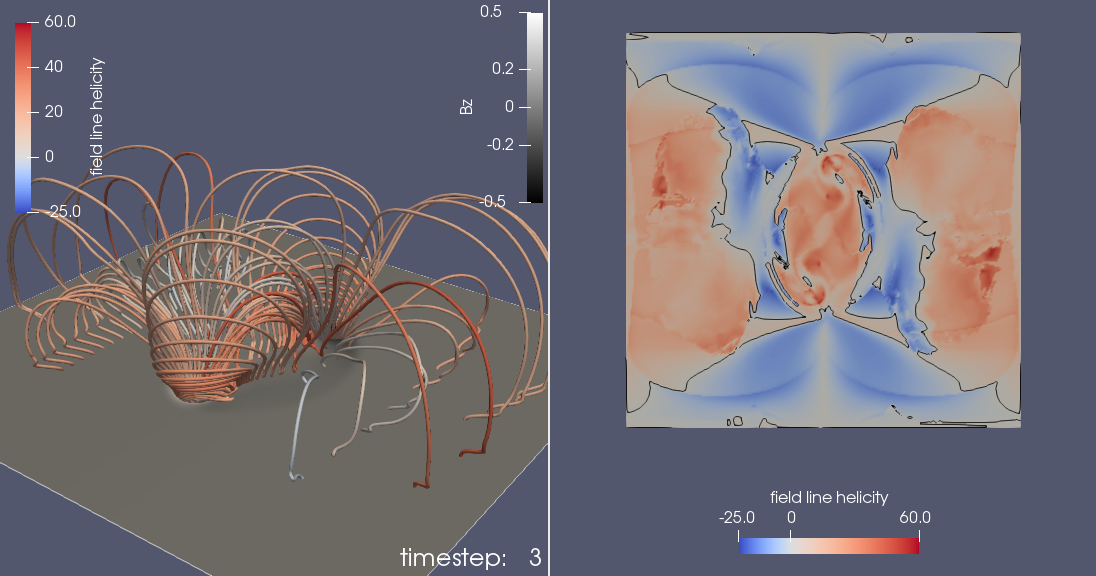}\\
\includegraphics[width=0.47\textwidth]{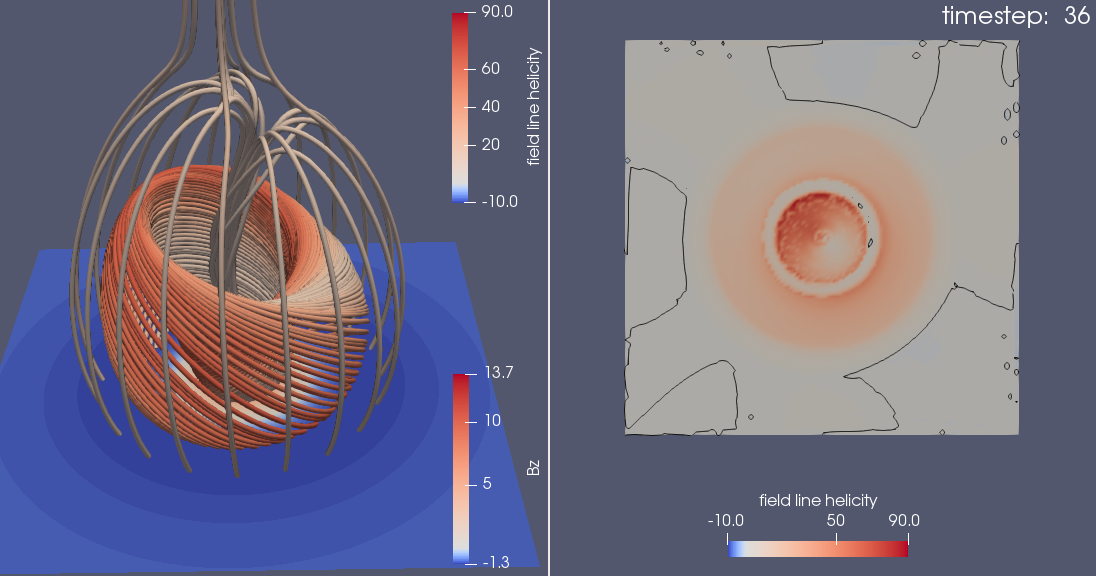}
\caption{Field line helicity morphology for the snapshot at $t=120$ of the non-eruptive emergence simulation (top), the snapshot at $t=110$ of the eruptive emergence simulation (middle), and the snapshot at $t=920$ of the jet formation simulation (bottom). The left plots show representative field lines colored by the value of their RMFLH overplotted on the photospheric map of $B_z$. In the top two plots, field lines are drawn from the contours $B_z=\pm 0.3$ and the lines at $x=\pm50$, while in the bottom plot from the contour $B_z=5$ and the circle of radius $r=3.4$ centered on the image. The right plots show the respective 2D maps of RMFLH on the photospheric plane.}
\label{compfig2}
\end{figure}

The main advantage of using RMFLH is that in addition to the value of helicity we can also infer its spatial distribution. We show that in two ways in Fig.~\ref{compfig2} using a snapshot from each MHD simulation. On the left column of Fig.~\ref{compfig2}, we present the 3D morphology of RMFLH by plotting representative field lines colour-coded according to the value of $\mathcal{A}_r^0$ for that field line, while the 2D distribution of $\mathcal{A}_r^0$ on the photospheric plane is displayed on the right column.

For the non-eruptive emergence simulation in the top panel of Fig.~\ref{compfig2}, and the shown snapshot at $t=120$, we notice that the emerging flux rope in the center of the image consists of field lines with high positive values of RMFLH. The flux rope is confined by the overlying arcade field which has mixed-signed, but mostly negative RMFLH of quite lower values, as can be seen in the 2D map of RMFLH on the right image.

For the eruptive emergence simulation and the snapshot at $t=110$ just before the onset of the eruption in the middle panel of Fig.~\ref{compfig2}, we notice a much more complex situation. The arcade field consists now mostly from positive RMFLH, again of lower values than the flux rope. Additionally, the flux rope itself has smaller values of RMFLH than in the non-eruptive case, confirming the smaller helicity budget shown in Fig.~\ref{compfig0}. 

\citet{pariat17} showed that the non-eruptive simulation had a larger positive helicity than the eruptive one. Using a toy model and the concept of ``self'' and ``mutual'' helicities, they explain these differences by the fact that the emerging flux rope, while having the same positive helicity in both cases, was globally sharing positive helicity with the surrounding arcade field in the non-eruptive case and negative helicity in the eruptive one. Within that toy model, the eruptive simulation would possess opposite signs of helicity, i.e., positive RMFLH in the field lines of the flux rope and negative RMFLH in the field lines of the surrounding arcade field. On the contrary, the non-eruptive simulation would have a distribution of RMFLH uniform in sign, positive for both the flux rope and the arcade field.

The results obtained here with RMFLH present a different view. Both eruptive and non-eruptive simulations present positive and negative RMFLH in the arcade field surrounding the flux rope. The spatial distribution of negative and positive RMFLH in that arcade field is more complex for the eruptive simulation than for the non-eruptive one. However, the main difference lies in the intensity of the positive and negative values of RMFLH. The eruptive case has a low total relative helicity mainly because all the domain possesses a weak RMFLH intensity. The non-eruptive case on the contrary, presents a high contrast of positively and negatively signed RMFLH. This further confirms that although ``self'' and ``mutual'' helicities are interesting theoretical tools, they are not directly relevant to relative magnetic helicity, and cannot be used to predict the distribution of RMFLH. The representation of a magnetic field as a collection of flux tubes is of course valid theoretically, and the relative magnetic helicity can then be expressed through sums over the self and mutual helicities under the specific gauge conditions described in \citet{dem06}. In the general case however, the approach of RMH with the self and mutual helicities is of limited practical use.

For the jet formation simulation, we show in the bottom panel of Fig.~\ref{compfig2} the snapshot at $t=920$ when the non-ideal effects take in. The twisting motions on the photosphere lead to the increase of RMFLH, and also of twist, in the central structure that later reconnects with the overlying jet-like feature. The values of RMFLH are predominantly positive in accordance with the results of \citet{pad09} and \citet{linan18}.

We should mention at this point that all these conclusions depend on the chosen gauges for the vector potentials. If RMFLH was expressed in a different gauge, its magnitude and even its sign distribution could be quite different. A preliminary testing of this possibility showed that this is not the case, and that at least in intense-RMFLH areas, the sign and magnitude of RMFLH are robust to the choice of the gauge. Some evidence in favour of this result can also be found in \citet{yeates18}.

We also note that an alternative to removing the gauge dependency of RMFLH is to specify the gauge explicitly, although this would correspond to a redefinition of relative magnetic helicity less general than the one of \citet{fa85}. Nevertheless, the resulting FLH can then be associated with a physically meaningful quantity. Such an example is given by the so-called winding gauge \citep{prior14}, which is a modified, planar Coulomb gauge. The respective FLH, that is termed the flux function in \citet{prior14}, expresses the average pairwise winding between the magnetic field lines. Nonetheless, it relies on a specific setup of the magnetic field,
where the lateral boundary consists only from flux surfaces, and cannot extend to the more general case considered here, where field lines starting from one side of the boundary can end in any side, lateral or not.

\subsection{Application to extrapolation data}
\label{sect:application3}

\begin{figure*}[h]
\centering
\includegraphics[width=0.42\textwidth]{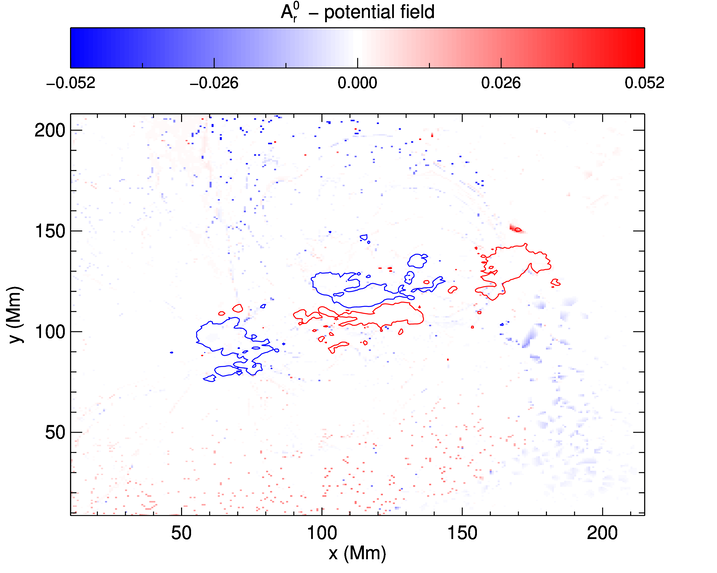}%
\includegraphics[width=0.42\textwidth]{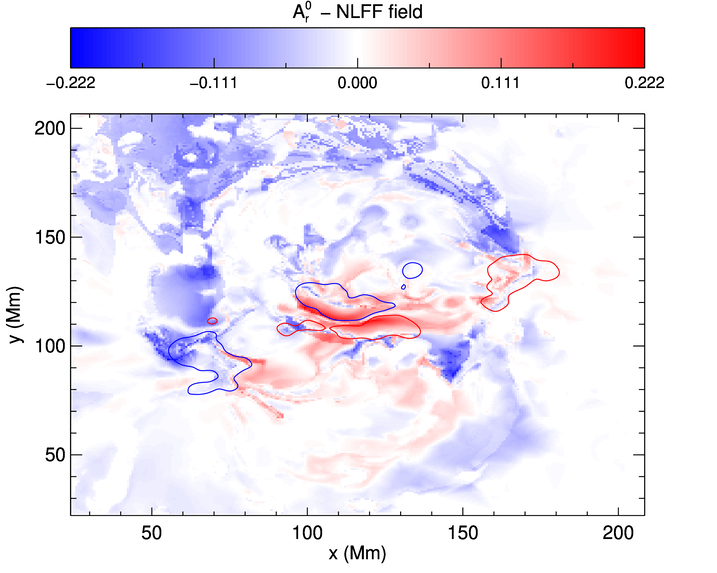}\\[0pt]
\includegraphics[width=0.42\textwidth]{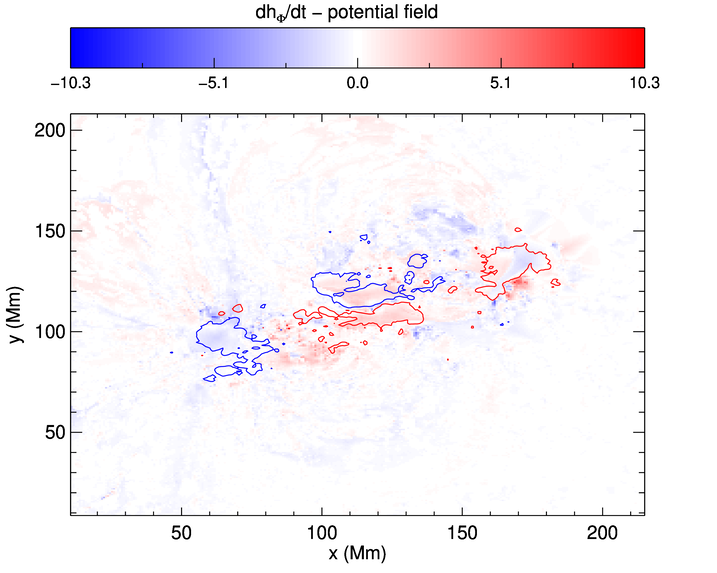}%
\includegraphics[width=0.42\textwidth]{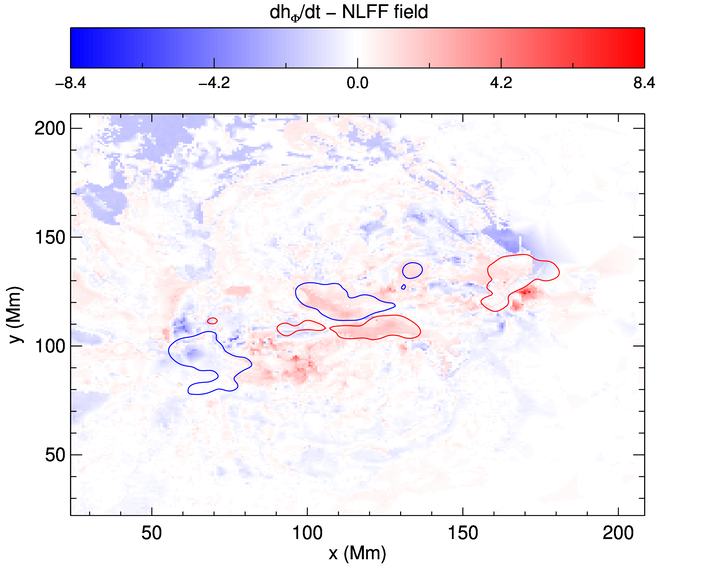}
\caption{Comparison of the 2D photospheric distribution of relative magnetic field line helicity (top) with the connectivity-based helicity flux density (bottom), for the potential (left), and the NLFF field extrapolations (right). The blue (red) contours correspond to $B_z=-500\,\mathrm{G}$ ($B_z=500\,\mathrm{G}$), and the colorbar units are $10^{22}\,\mathrm{Mx}$ for RMFLH and $10^{16}\,\mathrm{Mx} \,\mathrm{s}^{-1}$ for ${\rm d}h_\Phi/{\rm d}t$.}
\label{compfig3}
\end{figure*}

Here we use the extrapolation data presented in Sect.~\ref{sect:data2} to compare the 2D morphology of RMFLH with that of helicity flux density, ${\rm d}h_\Phi/{\rm d}t$, \citep{dalmasse18}. For the comparison to be meaningful we use the same footpoint locations in both calculations, i.e., only one method for tracing field lines is used, the one described in Sect.~\ref{sect:application1}. Finally, only field lines starting and closing on the photosphere are considered, since ${\rm d}h_\Phi/{\rm d}t$ is defined for closed field lines. The quantities $\mathcal{A}_r^0$, as described in Sect.~\ref{sect:application1}, and ${\rm d}h_\Phi/{\rm d}t$, as described in \citet{dalmasse14}, are then computed for this set of footpoints. The morphology of the two quantities on the photospheric plane for the two considered extrapolations is shown in Fig.~\ref{compfig3}. The comments in Sect.~\ref{sect:application2} about the gauge dependence of RMFLH also apply here, where, moreover, the two quantities examined are computed under different gauges.

For the extrapolation using the potential field assumption, by definition relative magnetic helicity should be null since the potential field is used as a reference field. We note that indeed RMFLH is insignificant in that case as shown in the top-left panel of Fig.~\ref{compfig3}. This is another indication that the RMFLH method is consistent. In practice of course helicity is not null, but very low, because the potential field computed by our method differs slightly from the extrapolated one. The helicity budget we find with the volume method is $H_1=-7.1\,10^{39}\,\mathrm{Mx}^2$, while with the RMFLH method it is $H_2=-7.5\,10^{39}\,\mathrm{Mx}^2$, and so the two methods show an agreement of $6\%$. These values may indicate the noise level that the helicity method can estimate.

Helicity flux density on the other hand is not necessarily negligible even when the magnetic field is potential, with strong helicity fluxes possibly  injecting helicity in some regions and removing it in others. The helicity flux density calculations associated with Fig.~\ref{compfig3} were checked against those of \citet{dalmasse18}, and they were found to be consistent, despite the different footpoints and field line integration methods considered in the two cases. This is also evident from the qualitative agreement of the 2D helicity flux density map displayed in the bottom left panel of Fig.~\ref{compfig3}, with the one presented in the top right Fig.~3 of \citet{dalmasse18} for a similar quantity. This further confirms the results of \citet{dalmasse18} that the computation of the helicity flux is robust to the selection of the starting field-line footpoints.

In the case of the NLFF extrapolation there is an intense relative helicity  budget in the domain, three orders of magnitude larger than in the case of the potential field for the RMFLH method. The value obtained by the volume method is $H_1=9.4\,10^{41}\,\mathrm{Mx}^2$, while the RMFLH method gives $H_2=1.3\,10^{42}\,\mathrm{Mx}^2$. The agreement is lower in this case, with the RMFLH method overestimating helicity by $\sim 30\%$. This level of agreement can still be considered acceptable given that the NLFF field is of poorer quality in terms of solenoidality as compared to the MHD ones, and solenoidal errors are the most important source of errors in helicity computations \citep[see Sect.~7 in][]{valori16}.

Although ${\rm d}h_\Phi/{\rm d}t$ is a flux, and thus a quantity of a different nature than RMFLH, we see in Fig.~\ref{compfig3} (right panels) that the overall distributions of the two quantities agree quite well in the case of the NLFF field extrapolation. As an example, in the two central contours RMFLH and ${\rm d}h_\Phi/{\rm d}t$ both exhibit the same sign and both are important, and so the positive helicity is increasing by the injection of more helicity there. There are regions where the two quantities are quite different of course, such as the negative RMFLH area centered at $(x,y)=(60,120)$ in Fig.~\ref{compfig3}, where ${\rm d}h_\Phi/{\rm d}t$ is mostly positive and not as strong. Nevertheless, the overall agreement of the two quantities is quite good.

If the helicity flux injection had been found to be strongly different from the volume helicity distribution, it would have implied that either at this particular time the helicity flux injection was very peculiar (something possible of course, e.g., for a twisted magnetic field being untwisted), or that the helicity flux method and the RMFLH were not carrying consistent information. However, these comparative results of the relative helicity distribution and relative helicity injection draws a coherent picture of the helicity properties in AR 11158, with helicity of a given sign being injected in regions having the same sign of helicity. This behaviour reasonably fits with what is expected to occur in solar active regions, helicity flux injection being relatively homogeneous and consistent in time. This is a first demonstration, that the common use of helicity flux and RMFLH can permit to deeper study the dynamics of helicity in solar active regions.

%%%%%%%%%%%%%%%%%%%%%%%%%%%%%%%%%%%%%%%%%%%%%%%%%%%%%%%%%%%%%
\section{Discussion}
\label{sect:discussion}

In this paper we defined the concept of relative magnetic field line helicity (RMFLH) and we examined some of its properties. We presented the mathematical derivation of RMFLH starting from the definition of RMH, without any assumption on the gauges of the vector potentials. A similar derivation was followed for the two gauge-invariant components of RMH, the non-potential and volume-threading ones. This decomposition of RMH has been shown to provide very interesting physical insights in the stability of magnetic structures \citep{pariat15,linan18}. All the derivations were based on the sole assumption that the magnetic field can be represented with flux tubes, a reasonable assumption for relatively smooth magnetic fields, and that these flux tubes start and end on the boundary of the volume, i.e., they do not form closed loops inside the volume.

The developed RMFLH does not pose any restriction on the vector potentials' gauge and so it can use the simplest possible gauge and/or the most economic in computational resources. By exploiting this flexibility in the choice of the gauge, we presented a fast and robust method for calculating RMFLH. We showed that the developed RMFLH can reproduce the system's relative magnetic helicity budget to high accuracy by using three different MHD simulations, and comparing with the results of a finite volume method of known good performance. 

A number of possible gauge combinations that simplify the resulting RMFLHs at the cost of more restrictive conditions for the computations of the vector potentials were also given. Within this framework, the recently proposed relative FLH of \citet{yeates18} is derived as a particular choice. 

Computations of RMFLH are performed for each snapshot separately as in any volume method. The computational cost for RMFLH is larger than a finite-volume method with the same gauge assumptions however, since it involves additionally the tracing of the field lines of two magnetic fields, original and potential. The information gained is also higher though, since RMFLH indicates the locations where helicity is more important, instead of providing a single value for the entire considered domain.

In the MHD simulations examined in this work for example, RMFLH reaffirmed the difference in the helicity budgets between the non-eruptive and the eruptive case, while in the jet case, more photospheric twisting lead to localized higher positive values for RMFLH in the region of twist injection. These results, and most notably the sign of RMFLH, are a function of the gauge of course, since RMFLH is a gauge-dependent quantity. Preliminary testing with different gauge combinations shows that locations of higher absolute values of RMFLH are unaffected by the gauge choice. This is also supported by the results of \citet[][Figs.~4 and 7]{yeates18}. More detailed examination of this dependence goes beyond the aim of this paper which is to introduce the quantity RMFLH. However, a representation and/or visualization of RMFLH that is less-, or non-related from the gauge is something worth exploring in the future.

In a first application of the developed RMFLH method to solar observations, relative magnetic field line helicity was shown to relate to helicity flux density. In a NLFF reconstruction of the magnetic field of a solar active region the two quantities showed similar overall photospheric spatial distribution, although local differences were also observed. The examination of RMFLH in similar future studies may thus provide additional information on the physical conditions of solar active regions.

Consequently, the concept of relative magnetic field line helicity has important potential for use in plasma applications. Field line helicity seems as a promising tool to apply especially to situations such as the flux ropes of coronal mass ejections \citep{green18}, where magnetic helicity is known to play an important role.

%%%%%%%%%%%%%%%%%%%%%%%%%%%%%%%%%%%%%%%%%%%%%%%%%%%%%%%%%%%%%
\begin{acknowledgements}
EP and KM acknowledge the support of the French Agence Nationale pour la Recherche through the HELISOL project, contract n$^\mathrm{o}$ ANR-15-CE31-0001. GV acknowledges the support of the Leverhulme Trust Research Project Grant 2014-051. This work and KD were supported by the French Centre National d'Etudes Spatiales. The authors acknowledge access to HPC resources of CINES under the allocations 2016–046331, A0010406331 and A0040406331 made by GENCI (Grand \'{E}quipement National de Calcul Intensif). We also appreciate the support of the International Space Science Institute for the international team ``Magnetic Helicity in Astrophysical Plasmas''.
\end{acknowledgements}

%%%%%%%%%%%%%%%%%%%%%%%%%%%%%%%%%%%%%%%%%%%%%%%%%%%%%%%%%%%%%
\bibliographystyle{aa}
\bibliography{refs}

%%%%%%%%%%%%%%%%%%%%%%%%%%%%%%%%%%%%%%%%%%%%%%%%%%%%%%%%%%%%%
\begin{appendix}
\label{appdx}

\section{Dependence of RMFLH on the choice of the footpoints}

As mentioned in Sect.~\ref{sect:application1} the default method for choosing the footpoint locations is by the grid points (GP) of the given magnetic field. Here we give an alternative method that assigns more footpoints to areas of stronger magnetic field strength. We refer to this method as the flux-based (FL) method.

\begin{figure}[h]
\centering
\includegraphics[width=0.35\textwidth]{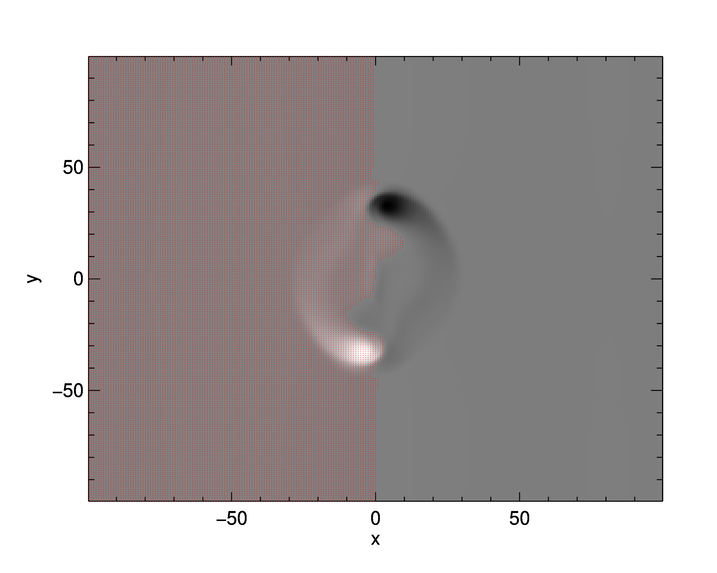}\\
\includegraphics[width=0.35\textwidth]{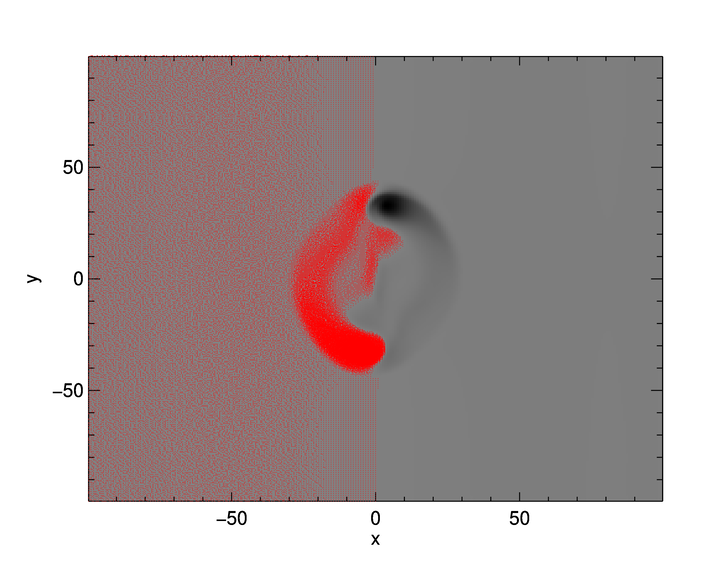}
\caption{Footpoint locations for the positive polarity of the snapshot at $t=120$ of the non-eruptive emergence simulation, as selected by the grid-point (top), and the flux-based (bottom) methods. Each footpoint is marked with a red dot, and they are overplotted on the $B_z$ photospheric map.}
\label{compftpts}
\end{figure}

In this, the magnetogram under study is first divided into weak and intense magnetic field areas. Typically, the division corresponds to the absolute value of $B_z$ above which $90\%$ of the flux in the magnetogram is taken into account. In the weak-field areas, a footpoint is assigned to each pixel as in the GP method. In the intense-field areas however, the footpoints assigned per pixel are proportional to the flux in that pixel. The lateral and top boundaries where the magnetic field is normally much weaker are treated as weak-field areas. We note also that this procedure results in the footpoints having similar values of magnetic flux, which is determined from the threshold value of $B_z$. As an example, we show in Fig.~\ref{compftpts} the photospheric footpoints obtained with the two methods for the snapshot at $t=120$ of the non-eruptive emergence simulation, and the positive-polarity boundary. The number of footpoints in the GP case are $\sim 27\,000$, while in the FL case almost three times that, $\sim 77\,000$.

\begin{figure}[h]
\centering
\includegraphics[width=0.44\textwidth]{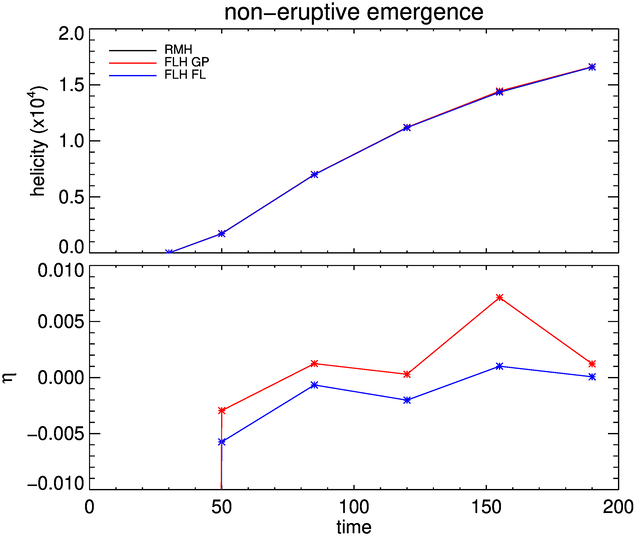}
\caption{Evolution of relative magnetic helicity as calculated by the volume method (with Eq.~(\ref{helr}), black curve) and by the field line method (with Eq.~(\ref{flhhel})) and footpoints selected by GP (red curve), and FL (blue curve) methods (top), and of their relative difference $\eta$ (bottom), for the non-eruptive emergence simulation.}
\label{compftptt}
\end{figure}

\begin{figure}[h]
\centering
\includegraphics[width=0.37\textwidth]{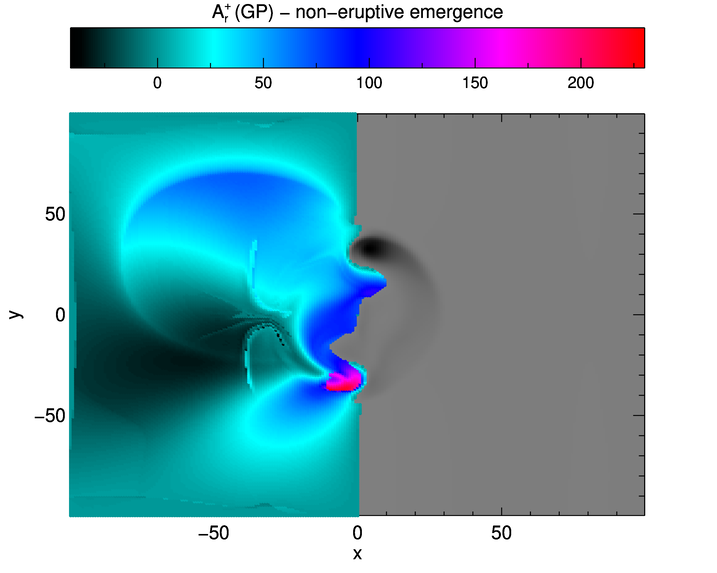}\\
\includegraphics[width=0.37\textwidth]{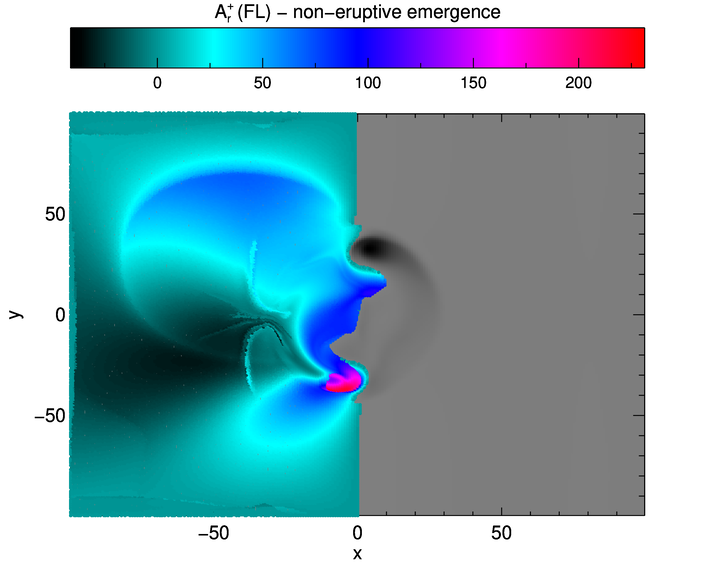}\\
\includegraphics[width=0.37\textwidth]{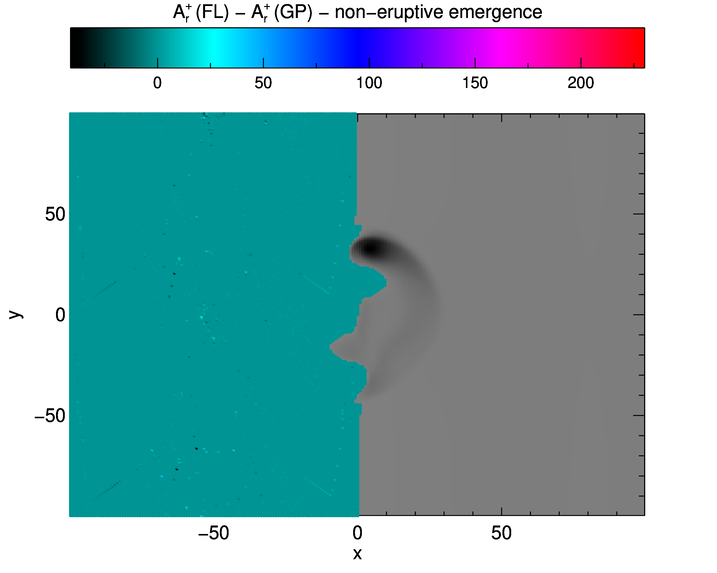}
\caption{2D morphology of the RMFLH $\mathcal{A}_r^+$ for the footpoints shown in the GP (top) and FL (middle) cases of Fig.~\ref{compftpts}, and their difference (bottom), overplotted on the $B_z$ photospheric map.}
\label{compftpty}
\end{figure}

The resulting helicity budget using these footpoints for the same simulation is shown in Fig.~\ref{compftptt}. For simplicity we have only considered the footpoints starting from the photospheric boundary. We notice in the bottom panel of Fig.~\ref{compftptt} that the relative (to the volume method) differences are very small, $\leq 1\%$ for both cases. This means that the GP method that is mainly used in this paper works equally well as the more accurate FL one, and so the choice of the former as more convenient, is justified. Moreover, the differences in the form of the RMFLH $\mathcal{A}_r^+$ computed with the two methods are very small, as is shown in Fig.~\ref{compftpty}.

\begin{figure}[h]
\centering
\includegraphics[width=0.37\textwidth]{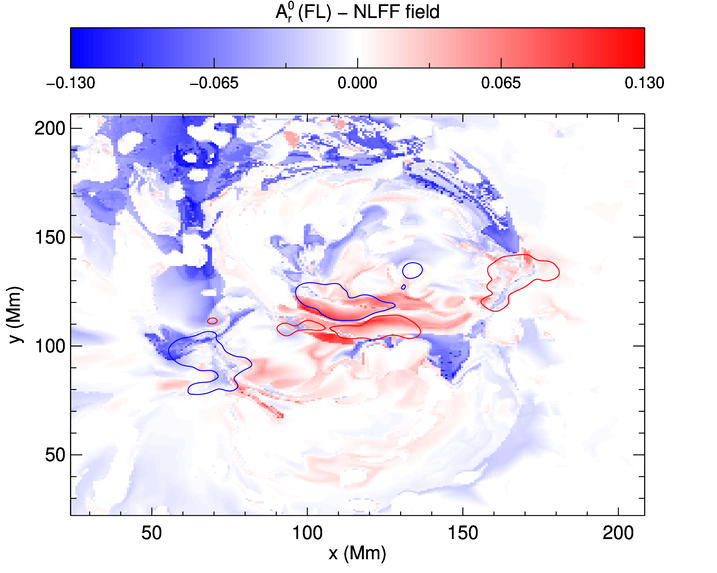}\\
\includegraphics[width=0.37\textwidth]{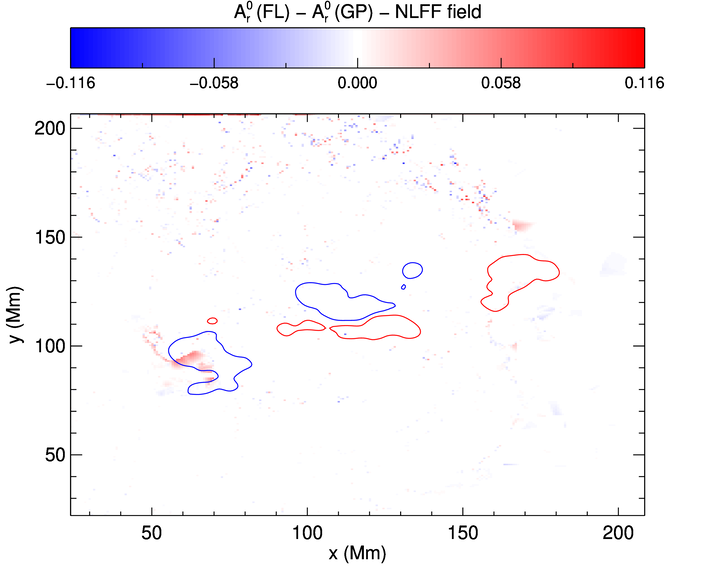}
\caption{Same as the top right panel of Fig.~\ref{compfig3} but for the footpoints chosen with the FL method (top), and difference of the two figures (bottom).}
\label{compftptz}
\end{figure}

To further examine the differences between the GP- and FL-based methods for choosing footpoints, we use the much more complex NLFF field reconstruction of AR 11158. In the top panel of Fig.~\ref{compftptz} we show the form of the RMFLH $\mathcal{A}_r^0$ computed with the footpoints given by the FL method for the threshold $B_z\simeq 2\,\mathrm{G}$ between weak and intense magnetic field areas. The differences with the top right Fig.~\ref{compfig3} where the footpoints are given by the GP method are very small, as the bottom panel of Fig.~\ref{compftptz} confirms.

\end{appendix}

\end{document}